Short Paper*

# Education in the Digital World: From the Lens of Millennial Learners

Jonelle Angelo S. Cenita
Richwell Colleges, Incorporated, Philippines
cenita1994@gmail.com
(corresponding author)

Zyra R. De Guzman
Richwell Colleges, Incorporated, Philippines
deguzman.zr@pnu.edu.ph

## Abstract

*Purpose* – The objective of this study is to determine Education in the Digital World: from the lens of millennial learners. This also identifies the cybergogical implications of the issue with digital education as seen through the lens of the outlier.

*Method* – This study uses a mixed-methods sequential explanatory design. A quantitative method was employed during the first phase and the instruments of the study were distributed using google forms. The survey received a total of 85 responses and the results were analyzed using descriptive methods. Following up with a qualitative method, during the second phase the outliers were interviewed, and the results were analyzed using thematic analysis. The results of the mixed methods were interpreted in the form of cybergogical implications.

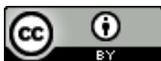

*Results* – The digital education from the lens of millennial learners in terms of the Benefits of E-Learning and Students' Perceptions of E-Learning received an overall mean of 3.68 which was verbally interpreted as "Highly acceptable". The results reveal that millennial learners' perceptions of digital education are influenced by the convenience in time and location, the fruit of collaboration using online interaction, the skills and knowledge they will acquire using digital resources, and the capability of improving themselves for the future.

*Conclusion* – E-learning significantly improves the quality of the learning and teaching process. Millennial learners with different learning styles and speeds were addressed by its usability and portability features. Millennial learners were able to adopt and learned how to use e-learning. Also, since it is self-paced learning, it allows them to study on their own time and schedule since e-learning can be accessed anytime and anywhere. However, the technological resources of the learners should be considered in the implementation of e-learning.

*Recommendations* – The utilization of e-learning as a medium of instruction for millennial learners.

*Research Implications* – The findings can be used by the institution to create guidelines, procedures, and policies for successfully implementing digital education or E-Learning.

*Keywords* – Education, Technology, Online Platform, Students perceptions, Millennial learners

## INTRODUCTION

Online education has rapidly increased over the years, and the increased accessibility of the internet created an opportunity for non-traditional education or E-Learning (Li & Irby, 2008). Before the Covid-19 pandemic, learners already utilized the internet for their personal development. According to Slamti and Ajrouh (2019), Millennial learners use free online courses for their personal development as complementary to their studies. Research by Hass and Joseph (2018) found that the overall respondents of the study have a neutral perception of online courses, while the majority of the respondents would likely take online courses. The study by Seale et al. (2021), found out how technology-enabled the students to achieve their academic potential. Furthermore,

Ametova and Mustafoeva (2020), state that education can be practiced anywhere, anytime, and can be accessed using various electronic devices such as mobile phones.

On 30th January 2020, The World Health Organization (WHO) declared COVID-19 as a global public health emergency of international concern and a pandemic on 11th March 2020 (Cucinotta & Vanelli, 2020). According to Khan et al. (2020), Muthuprasad et al. (2021), and Toquero (2020), Educational institutions across the world have been so vastly affected that they are forced to shut down due to the outbreak of the COVID-19 pandemic and jeopardized the academic calendar. Khalil et al. (2020) and Paudel (2021) state that the closure of educational activities due to the ongoing COVID-19 pandemic has led many educational institutions to forcefully shift the mode of education from face-to-face to online learning. Previous studies (Sato, 2020; Mclaughlin et al., 2020) state that education should not stop despite pandemics and discuss the importance of online learning platforms to continue providing education for students.

Williams et al. (2020) and Khan et al. (2020) conducted a study on students' perspectives toward e-learning or online education. The results indicate that students' positive perceptions in an online class may equal or surpass traditional education. According to Paudel (2021), online education can be an alternative means to traditional education. Hence, to ensure effectiveness and a successful learning process, a blended approach is recommended. Thus, Alshehri (2017), Mitchell (2020), and Dogar et al. (2020) stated that students' inability to utilize digital infrastructures such as high-speed internet and electronic devices is the major constraint to learning using online education. Moreover, distractions significantly affect students' learning and performance (Schmidt, 2020).

The main objective of this study is to determine Education in the Digital World: from the lens of Millennial Learners. This also summarizes the problem encountered in digital education from the outlier's lens and identifies the cybergogical implications. The findings of the study can use by the institution to craft guidelines, procedures, and policies to implement digital education or E-Learning successfully.

## METHODOLOGY

This study uses a mixed-methods, explanatory design. Explanatory Design begins with a Quantitative approach follow up by a Qualitative approach to understand the problem in depth and determine the root cause of a certain situation and fill gaps in missing information (Figure 1).

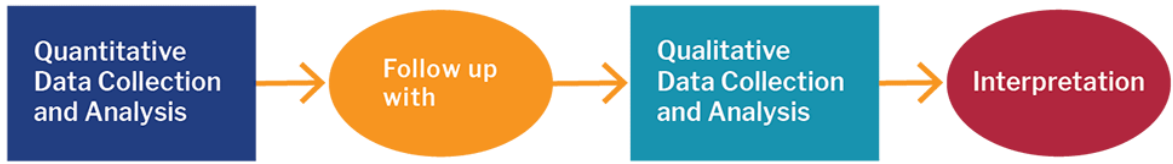

*Figure 1.* Explanatory Design

## *Quantitative*

During the first phase, the researcher distributed the online survey instruments of the study via Google Forms and rated them using the five-point Likert scale. The online survey received a total of 85 responses and the gathered results were statistically computed using a weighted mean and interpreted using descriptive analysis.

## *Qualitative*

During the second phase, the researcher interviewed the outlier and the results of the interview with the two outliers were analyzed using thematic analysis. The outlier came from the weighted mean of the learner's responses from the distributed instruments of the study producing an outlier. The results of the mixed methods were interpreted in the form of cybergogical implications.

## *Respondents of the study*

The study respondents are the students from Technical Vocational students of Richwell College, Inc. The respondent was selected using purposive non-probability sampling as they are the ones who actively use E-Learning platforms as the medium of instruction.

## *Instruments of the Study*

The instruments of the study were developed by Khan et al. (2020) (Table 1). The researcher validated the instruments using Cronbach's alpha and received 0.979, which indicates excellent internal consistency allowing for further analysis.

Table 1. Benefits of E-Learning and Students' Perceptions of E-learning Online Survey Instruments

| Variables | Rating Scale | | | | |
|---|---|---|---|---|---|
| | 1 (NA) | 2 (LA) | 3 (MA) | 4 (HA) | 5 (VHA) |
| **Benefits of E-Learning** | | | | | |
| B.1 Flexibility in Time and Space | | | | | |
| B.2 Ease and quick sharing of educational material | | | | | |
| B.3 Improved collaboration and interactivity among students | | | | | |
| B.4 Access to higher education for all applicants | | | | | |
| B.5 Possibility of working with e-learning | | | | | |
| B.6 Accommodates different types of learning styles | | | | | |
| B.7 Quick Feedback | | | | | |
| B.8 Wide and diverse interactions | | | | | |
| B.9 Access study resources effectively | | | | | |
| B. 10 Updated learning material | | | | | |
| **Perceived Usefulness of E-Learning** | | | | | |
| PU. 1 Studying through the e-learning model provides flexibility to the study at a time convenient to the learner. | | | | | |
| PU. 2 E-learning can enable people to study irrespective of where they are located in the world. | | | | | |
| PU. 3 There are technologies available to enable one to take tests and submit assignments electronically. | | | | | |
| PU. 4 There are electronic tools available to enable interactive communication between instructor and student without meeting face-to-face. | | | | | |
| **Perceived Self-Efficacy of Using E-Learning** | | | | | |
| PS. 1 I feel confident while using the e-learning system. | | | | | |
| PS. 2 I feel confident while operating e-learning functions. | | | | | |
| PS. 3 I feel confident while using online learning content. | | | | | |
| **Perceived Ease of Use of E-Learning** | | | | | |
| PE. 1 I believe e-learning platforms are user-friendly. | | | | | |
| PE. 2 It would be easy for me to find necessary information when using an e-learning platform. | | | | | |
| PE. 3 I believe that using e-learning services can simplify the learning process. | | | | | |
| PE. 4 The set-up of the e-learning service is compatible with the way I learn. | | | | | |
| **Behavioral Intention of Using E-Learning** | | | | | |
| BI. 1 I intend to use e-learning to assist my learning. | | | | | |
| BI. 2 I intend to use e-learning to get updated my subject knowledge with the latest amendments. | | | | | |
| BI. 3 I intend to use e-learning as an autonomous (free) learning tool. | | | | | |

The researcher uses Five-Point Likert scale-based statements ranging from Not Acceptable to Very Highly Acceptable (Table 2).

Table 2. Likert Scale

| Rating Scale | Range | Descriptive Interpretation |
|---|---|---|
| 5 | 4.5– 5.00 | Very Highly Acceptable |
| 4 | 3.5 – 4.49 | Highly Acceptable |
| 3 | 2.5 – 3.49 | Moderate Acceptable |
| 2 | 1.5 – 2.49 | Least Acceptable |
| 1 | 1.0 – 1.49 | Not Acceptable |

## RESULTS

### *Benefits of E-Learning and Students' Perceptions of E-learning*

Table 3 conveys that on average, the learners agreed on the 'benefits of E-learning" and "perceived usefulness of E-learning," "perceived self-efficacy of using E-learning," "perceived ease of use of E-learning,' and the "behavioral intention of using E-learning" based on the weighted mean of 3.68, 3.75, 3.61, 3.59, 3.74, respectively, which have a verbal description of "Highly Acceptable," The overall mean of 3.67 was computed at which was verbally interpreted as "Highly Acceptable."

Table 3. Students' perceptions

| Variable | Mean | Descriptive Interpretation |
|---|---|---|
| Benefits of E-Learning | 3.68 | Highly Acceptable |
| Perceived Usefulness of E-Learning | 3.75 | Highly Acceptable |
| Perceived Self-Efficacy of Using E-Learning | 3.61 | Highly Acceptable |
| Perceived Ease of Use of E-Learning | 3.59 | Highly Acceptable |
| Behavioral Intention of Using E-Learning | 3.74 | Highly Acceptable |
| **Mean** | **3.67** | **Highly Acceptable** |

Table 3 shows that "Perceived Usefulness of E-Learning" received the highest response rate of 3.75 with the descriptive interpretation of "highly accepted." This shows that the flexibility of e-learning allows students to study anytime and anywhere. "Behavioral Intention of Using E-Learning" received 3.74 with the descriptive interpretation of "highly accepted." It appears that students prefer to use e-learning to get their knowledge updated. "Benefits of E-Learning" received 3.68 with the descriptive interpretation of "highly accepted." This indicates that the benefits of e-learning give convenience to the students in terms of time, place, pace, and educational material. "Perceived Self-Efficacy of Using E-Learning" received 3.61 with the descriptive interpretation of "highly accepted." This expresses that students are confident in the e-learning approach and its contents. And lastly, "Perceived Ease of Use of E-Learning" received 3.61 with the descriptive interpretation of "highly accepted." This appears that students are compatible with the way they learn, and e-learning is easy to use and makes the learning process simple.

The digital education from the lens of millennial learners in terms of the Benefits of E-Learning and Students' Perceptions of E-Learning received an overall mean of 3.68 which was verbally interpreted as "Highly acceptable". The results reveal that millennial learners' perceptions of digital education are influenced by the convenience in time and location, the fruit of collaboration using online interaction, the skills and knowledge they will acquire using digital resources, and the capability of improving themselves for the future.

## *Digital Education from Outliers Lens*

Figure 2 illustrates the statistical computation using the weighted mean of the learner's responses. The illustration reveals that there is an outlier among the respondents and it shows that millennial students encounter problems during digital education.

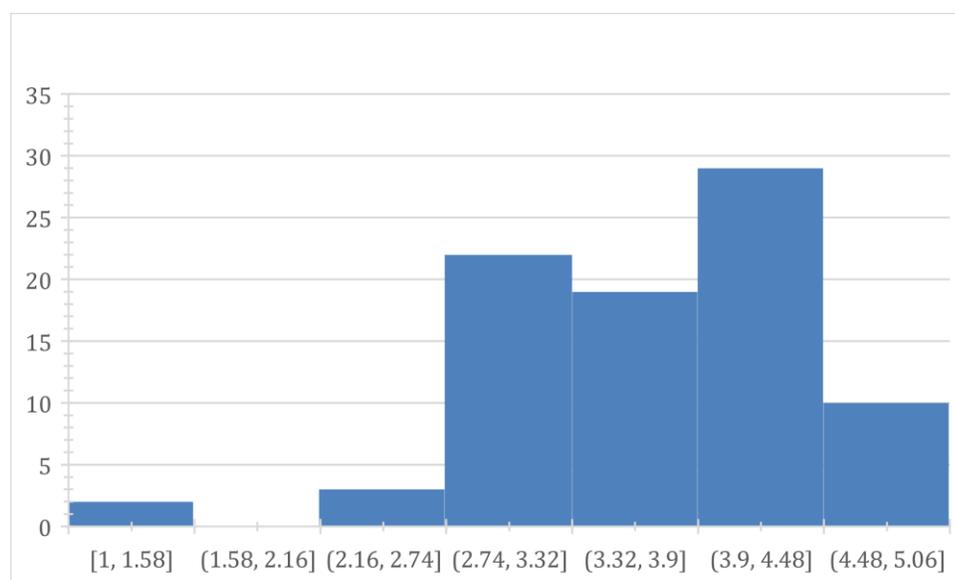

*Figure 2.* Outliers

Table 4. shows the result of the thematic analysis in the interview of two outliers. The outliers were interviewed by the researchers, who asked them to explain how they perceived each of the survey instruments' variables. The outlier responses were narrated,

interpreted, and organized thematically. This allows the researchers to explore, expand, and deepen their understanding of the data obtained from the online survey results.

Table 4. Outlier Perceptions

| Outlier Narrations | Interpretation |
|---|---|
| **Benefits of E-Learning**<br>**Student 1:**<br>"Para sakin po kasi mas ano po ako, mas natututo po ako kapag kaharap ko po yung teacher kapag nasa paligid kopo yung classmate ko po na momotitave po ako kasi po ngaun pong time parang pong hindi po kasi ako sanay mag isa nahihirapan po ako mag cope up sa bagong system"<br>"Ano po kami po kasi in terms of gadgets, hindi po kasi makakabili ng mga mas capable na gadget para sa online learning. Wala rin po kaming budget para sa internet connection"<br>"Sa time management po kasi habang nag aaral po kasi habang nag aaral po kasi kapag uutusan ng magulang, parang nawawala po yung focus sa pag aaral hindi tulad kapag nasa skwelahan kalang hindi kalang nakafocus"<br>"Okay naman sakin yung e learning sa pagsasagot po kami po kasi lms ang problema lang po talaga kapag nasa bahay, nawawala yung focus sa pag aaral"<br><br>**Student 2:**<br>"Sakin po kasi hindi saken po kasi mas natututo po ako kapag teacher kopo nag sasabi nag explain"<br>"Hindi po 6 po kami mag kakapatid tapos nag sasalo lang po kami sa isang cell phone"<br>"Hindi po kasi kapag nasa bahay po ako na stress po ako kasi uunahin kuba sila o uunahin ko yung sakin"<br>"Nakaka stress, nakaka depress lahat na" | It may be deduced from the outliers' responses that they will be more likely to learn if the teacher facilitates the learning process physically. Also, the respondents do not have technological resources to access E-Learning and the respondents have too many household responsibilities to not focus on their education in their homes. |
| **Perceived Usefulness of E-Learning**<br>**Student 1:**<br>"Halimbawa po, kung meron po other engagement sa everyday life, kahit nasan ka pwede mo i access pinag aaralan mo. hindi tulad nasa bahay kalang ganun."<br>"Mas madali ma access kaya lang maraming studyante hindi afford yung gadget and internet connection"<br><br>**Student 2:**<br>"Parang mas okay po nasa skwelahan kasi face to face kasi 6 nga po kami kasi kahit module nila iisipin kupa"<br>"Madali naman posyang gamitin lalo na kung maganda load ang ginamit ko saka yung signal" | It may be deduced from the outliers' responses that they agree that E-Learning is helpful because it can be accessed anywhere. However, the respondents do not have the technological resources to access E-Learning. Nevertheless, one respondent still prefers Face-To-Face education due to too many household responsibilities. |

Table 4. Outlier Perceptions (cont.)

| Outlier Narrations | Interpretation |
|---|---|
| **Perceived Self-Efficacy of Using E-Learning**<br>**Student 1:**<br>"Mas madali po mag sagot hindi napo gagastos sa mga ballpen papel kasi kami sa lms dun na kami mag sasagot."<br><br>**Student 2:**<br>"Siguro po depende po siguro sa signal na meron kami" | It may be deduced from the outliers' responses that they are confident in using e-learning if technological resources are available. |
| **Perceived Ease of Use of E-Learning**<br>**Student 1:**<br>"Mas madali napo kasi kapag sinearch mo yung tanong nandun napo agad yung sagot"<br>"Para sakin po, yung learning process parang naging mas kumplikado, kagaya ng tulad kong sinabi mas okay sakin may nakakausap akong classmate kasi na momotivate ako."<br>"Para po sakin ngaun parang puro comply nalng ginagawa ko, wala na akong nakikitang outcome kumpara dati sa lahat na a-adopt ko ngaun hindi na po."<br><br>**Student 2:**<br>"Yes po nandun na lahat basta may load lang makikita."<br>"Mas pinadali kasi hindi kana mag iisip na hindi tulad sa eskwelahan kailanga mo ipaliwanag lahat"<br>"Mas convenient po yung face to face." | It may be deduced from the outliers' responses that they agreed that the e-learning platform is easy to use. However, the respondents do not have the technological resources to access E-Learning. Also, they prefer Face-To-Face learning because it is the most convenient for them. Nevertheless, one respondent expresses that the e-learning process is complicated and he/she was focused only on complying with his/her academic requirements rather than learning. |
| **Behavioral Intention of Using E-Learning**<br>**Student 1:**<br>"Nawalan po ako ng confident sa mga sinasagot ko kasi nag rely nalang ako sa internet, hindi tulad dati na galing ang information sa teacher"<br>"Hindi po kasi mas gusto ko parin po talaga na sama sama sa classroom nakikita ko mga teacher. Mahirap po para sakin kasi wala kapong makakapalitan ng idea, ayon po kasi nakasanayan namin"<br>"Hindi po lalo napo sa math sa english po, hindi tulad ng dinideliver ng teacher namin natutunan po namin agad"<br>**Student 2:**<br>"Hindi po tlaga ako agree sa e learning o online class"<br><br>"Lahat napo ata nagawa kuna, hindi po kasi ako mahilig mag tanong sa kaklase ko."<br><br>"Siguro po iniintindi ko nalang yung deadline para makapag comply" | It may be deduced from the outliers' responses that they prefer Face-To-Face education. Also, the respondents do not have confidence in their answers and only do it for compliance. Furthermore, one respondent expresses that he/she will learn more if the teacher facilitates the learning process physically. |

## *Cybergogical Implications*

### *Convenient and Flexible*

The respondents interpreted E-learning as flexible and convenient to use. Students can access their instructional materials using an E-learning platform and study anytime, anywhere using any digital device. Furthermore, E-Learning gives students the element of control over time, place, and pace of study.

### *Collaboration and Interaction*

These are available technology tools for the students and teachers to collaborate and interact with one another. This technology allows the students to (1) Independent learning such as learning the topic, submitting the activity/assignment, and accomplishing their test; (2) Communicate and share their ideas and opinions by using different online platforms/peer-based resources; and (3) received quick feedback to their teachers after completing a specific task.

### *Ease to Use Platform*

The respondents highly accepted the nontraditional method of education, E-Learning. The results illustrated that students were secure and confident in using E-learning because of its intuitive user interface. The E-Learning platform simplifies student knowledge acquisition.

### *Personalized learning path*

As online education supports the new curriculum and teaching strategies. E-Learning platforms can be modified according to the learner's convenience and make it easy for them to gather information. The survey results show that e-learning promoted learning autonomy and improved the students' knowledge after using e-learning platforms.

## DISCUSSION

Songkram et al. (2015), and Ametova and Mustafoeva (2020) stated that E-Learning is beneficial to students because it enables them to take their classes anywhere, anytime, using any form of digital device. Moreover, giving them more time and opportunity to work. It is supported by this study's finding that E-Learning offers convenience in terms of time and location.

El Haddioui and Khaldi (2012), stated that E-Leaners have different types and varieties of learning approaches. The findings of this study reveal that respondents have different learning styles that should be considered. Digital education should provide convenience in the learning process and gives an appropriate learning style for any type of student.

The studies (Dutta & Smita, 2020; Nichols, 2020; Dogar et al., 2020) revealed that students' inability to utilize digital infrastructure such as high-speed internet and electronic devices is the major constraint to continuing learning using online education. Songkram (2015) found out that Information Communication Technology (ICT) integration in e-learning is improving the quality of learning and teaching. It is supported by this study's findings that Millennials learners further develop their knowledge after using e-learning.

## CONCLUSIONS AND RECOMMENDATIONS

In conclusion, the Integration of technology such as E-learning significantly improves the quality of learning and teaching process as shown in the results of the study. Millennial learners with different learning styles and learning speeds were addressed by its usability and portability features. Because of its features millennial learners were able to adopt and learned how to use e-learning. Also, since it is self-paced learning, it allows millennial learners to study on their own time and schedule since e-learning can be accessed anytime and anywhere. However, the technological resources of the learners should be considered in the implementation of e-learning. The utilization of e-learning as a medium of instruction for millennial learners with technological resources.

## IMPLICATIONS

The study may be extremely useful to the organization as it will evaluate millennial learners' perceptions of online learning. The findings can be used by the institution to create guidelines, procedures, and policies for successfully implementing digital education or E-Learning.

## ACKNOWLEDGEMENT

The researchers want to express deep gratitude to the ICpEP. Sincere appreciation to the conference organizers for organizing this event and for including this paper in the conference souvenir program. The researchers would like to acknowledge and express gratitude to Richwell Colleges, Inc. The researchers are grateful for your kind encouragement and motivation to carry out this research.

# DECLARATIONS

## *Conflict of Interest*

The authors declare that there is no conflict of interest.

## *Informed Consent*

The purpose of the study was explained to each participant. Moreover, how the data were gathered, put to use, stored, and disposed of. The participants were also made aware that it was possible they would be contacted for an interview to explain their responses. All survey participants received a reference number after completing the form, but they all remained anonymous.

## *Ethics Approval*

The institution does not have a research ethics committee, but the research department of the institution approved and monitored the study from the beginning to the end.

# REFERENCES

Alshehri, A. A. (2017). The Perception created of online homework by high school student, their teacher and parents in Saudi Arabia. *Journal of Education and Practice, 8*(13), 85-100.

Ametova, O. R., & Mustafoeva, N. I. (2020). The benefits and drawbacks of online education for law students in higher educational institutions. *ISJ Theoretical & Applied Science, 12*(92), 61-63.

Cucinotta, D., & Vanelli, M. (2020). WHO declares COVID-19 a pandemic. *Acta bio medica: Atenei parmensis, 91*(1), 157.

Dogar, A. A., Shah, I., Ali, S. W., & Ijaz, A. (2020). Constraints to online teaching in institutes of higher education during pandemic COVID-19: A case study of CUI, Abbottabad Pakistan. *Revista Romaneasca Pentru Educatie Multidimensionala, 12*(2Sup1), 12-24.

Dutta, S., & Smita, M. K. (2020). The impact of COVID-19 pandemic on tertiary education in Bangladesh: students' perspectives. *Open Journal of Social Sciences, 8*(09), 53.

El Haddioui, I., & Khaldi, M. (2012). Learner behavior analysis on an online learning platform. *International Journal of Emerging Technologies in Learning (iJET), 7*(2), 22-25.

Hass, A., & Joseph, M. (2018). Investigating different options in course delivery–traditional vs online: is there another option?. *The International Journal of Information and Learning Technology, 35*(4), 230-239.

Khalil, R., Mansour, A. E., Fadda, W. A., Almisnid, K., Aldamegh, M., Al-Nafeesah, Alkhalifah, A. & Al-Wutayd, O. (2020). The sudden transition to synchronized online learning during the COVID-19 pandemic in Saudi Arabia: a qualitative study exploring medical students' perspectives. *BMC medical education, 20*, 1-10.

Khan, M. A., Nabi, M. K., Khojah, M., & Tahir, M. (2020). Students' perception towards e-learning during COVID-19 pandemic in India: An empirical study. *Sustainability, 13*(1), 57.

Li, C. S., & Irby, B. (2008). An overview of online education: Attractiveness, benefits, challenges, concerns and recommendations. *College Student Journal, 42*(2), 449-459.

Mclaughlin, H., Scholar, H., & Teater, B. (2020). Social work education in a global pandemic: Strategies, reflections, and challenges. *Social Work Education, 39*(8), 975-982.

Mitchell, C. (2020). *Serving special needs students during COVID-19: A rural educator's story*. Education Week.

Muthuprasad, T., Aiswarya, S., Aditya, K. S., & Jha, G. K. (2021). Students' perception and preference for online education in India during COVID-19 pandemic. Social sciences & humanities open, 3(1), 100101.

Nichols, B. E. (2020). Equity in music education: Access to learning during the pandemic and beyond. *Music Educators Journal, 107*(1), 68-70.

Paudel, P. (2021). Online education: Benefits, challenges and strategies during and after COVID-19 in higher education. *International Journal on Studies in Education, 3*(2), 70-85.

Sato, H. (2020). Educational responses to the pandemic in Japan: Primary and secondary education policy issues. *International Studies in Educational Administration, 48*(2), 64-69.

Schmidt, S. J. (2020). Distracted learning: Big problem and golden opportunity. *Journal of Food Science Education, 19*(4), 278-291.

Seale, J., Colwell, C., Coughlan, T., Heiman, T., Kaspi-Tsahor, D., & Olenik-Shemesh, D. (2021). 'Dreaming in colour': disabled higher education students' perspectives on improving design practices that would enable them to benefit from their use of technologies. *Education and Information Technologies, 26*, 1687-1719.

Slamti, K., & Ajrouh, L. (2019). Online graduate degrees: Perceptions of Moroccan University students. In part of the Multi Conference on Computer Science and Information Systems 2019 (p. 257).

Songkram, N. (2015). E-learning system in virtual learning environment to develop creative thinking for learners in higher education. *Procedia-Social and Behavioral Sciences, 174*, 674-679.

Songkram, N., Khlaisang, J., Puthaseranee, B., & Likhitdamrongkiat, M. (2015). E-learning system to enhance cognitive skills for learners in higher education. *Procedia-Social and Behavioral Sciences, 174*, 667-673.

Toquero, C. M. (2020). Challenges and opportunities for higher education amid the COVID-19 pandemic: The Philippine context. *Pedagogical Research, 5 (4), em0063.*

Williams, L., Martinasek, M., Carone, K., & Sanders, S. (2020). High school students' perceptions of traditional and online health and Physical Education courses. Journal of School Health, 90(3), 234-244.

**Author's Biography**

Jonelle Angelo S. Cenita is a college instructor at Richwell Colleges, Incorporated. He is currently enrolled in the program of Doctor of Information Technology at La Consolacion University Philippines and is a graduate of Master of Science in Information technology and Bachelor of Science in Information Technology.

Zyra R. De Guzman is a college instructor at Richwell Colleges, Incorporated. She is currently enrolled in the program of Master of Arts in Language Education Major in Filipino at Philippine Normal University and a graduate of Bachelor of Secondary Education Major in Filipino and a licensed professional teacher.